\documentclass[floatfix,nofootinbib,showpacs,showkeys,preprintnumbers,eqsecnum]{revtex4}
\usepackage{dcolumn}%
\usepackage{bm}%
\usepackage{fullpage}
\usepackage[centertags]{amsmath}
\allowdisplaybreaks[1]
\usepackage{amsbsy}
\usepackage{amsfonts}
\usepackage{amssymb}
\usepackage{eufrak}
\usepackage[dvips]{graphicx}
\begin{document}
\preprint{hep-ph/0209169}
\title{CP Violation in Bilarge Lepton Mixing}
\author{Carlo Giunti}
\email{giunti@to.infn.it}
\homepage{http://www.to.infn.it/~giunti}
\affiliation{
INFN, Sezione di Torino, and Dipartimento di Fisica Teorica, Universit\`a di Torino,
Via P. Giuria 1, I--10125 Torino, Italy
}
\author{Morimitsu Tanimoto}
\email{tanimoto@muse.sc.niigata-u.ac.jp}
\affiliation{
Department of Physics, Niigata University,
Ikarashi 2-8050, 950-2181 Niigata, JAPAN
}
\date{22 October 2002}
\begin{abstract}
We propose a scheme of lepton mixing in which
the unitary matrix that diagonalizes
the neutrino mass matrix is bimaximal
and the deviation from bimaximal of the lepton mixing matrix
is due to the unitary matrix that diagonalizes
the charged-lepton mass matrix.
This matrix is assumed to be hierarchical,
like the quark mixing matrix.
It is shown that in general
it is possible to have a sizable value for $|U_{e3}|$
together with an effective
two-neutrino maximal mixing in solar neutrino experiments.
If the effective mixing in solar neutrino experiments
is less than maximal,
as indicated by current data,
$|U_{e3}|$ is bounded from below.
Furthermore,
in general the violation of CP could be relatively large.
\end{abstract}
\pacs{14.60.Pq, 14.60.Lm, 26.65.+t, 95.85.Ry}
\keywords{Neutrino Oscillations, Solar and Atmospheric Neutrinos, CP Violation}
\maketitle

\section{Introduction}
\label{Introduction}

Solar and atmospheric neutrino experiments found recently strong evidence
in favor of neutrino oscillations with large mixing.
The Super-Kamiokande experiment found
model independent evidence of
disappearance of atmospheric $\nu_\mu$'s
\cite{Fukuda:1998mi}
and
the SNO solar neutrino experiment
found model independent evidence of
$\nu_e\to\nu_\mu,\nu_\tau$
solar neutrino transitions \cite{Ahmad:2002jz}.
These model-independent
evidences are further supported by the results of the
K2K long-baseline experiment \cite{K2K-Nu2002},
of the Soudan 2 \cite{Allison:1999ms} and MACRO \cite{Ambrosio:2000qy}
atmospheric neutrino experiment
and
by the
Homestake \cite{Cleveland:1998nv},
GALLEX \cite{Hampel:1998xg},
SAGE \cite{astro-ph/0204245},
GNO \cite{Altmann:2000ft}
and
Super-Kamiokande \cite{Fukuda:2002pe}
solar neutrino experiments.

Three-neutrino mixing is the simplest and most natural
known explanation of the results of solar and atmospheric neutrino experiments
(see Ref.~\cite{Giunti-ICHEP02}, and
Ref.~\cite{Neutrino-Unbound} for an extensive and updated list of
references).
Taking into account also the negative results of
the CHOOZ long-baseline reactor neutrino experiment \cite{Apollonio:1999ae},
it turns out that
the lepton mixing matrix
must have a bilarge form,
\textit{i.e.} close to bimaximal
\cite{Vissani:1997pa,Barger:1998ta,Baltz:1998ey}.

In this paper we extend the scheme proposed
in Ref.~\cite{Giunti:2002ye},
in which bilarge lepton mixing is obtained
as a deviation from bimaximal mixing
due to the unitary matrix that diagonalizes the
charged lepton mass matrix.
With respect to Ref.~\cite{Giunti:2002ye},
in this paper we write the mixing matrix in a more general form\footnote{
After completion of this work,
we have been informed that a similar scheme has been discussed
by Zhi-Zhong Xing in Ref.~\cite{Xing:2001cx}.
},
adding phases that were omitted in
Ref.~\cite{Giunti:2002ye}.
As a consequence,
we will show that it is possible to have relatively large
CP violation in the lepton sector,
contrary to the results of
Ref.~\cite{Giunti:2002ye},
where CP violation was found to be very small.

In Section~\ref{Experimental Results}
we briefly review the relevant experimental results.
In Section~\ref{Lepton Mixing}
we describe our scheme, with the help of
Appendix~\ref{Parameterization and Rephasing of the Mixing Matrix}.
In Section~\ref{Phenomenology}
we discuss the phenomenological consequences of our scheme.
Conclusions are drawn in Section~\ref{Conclusions}.

\section{Experimental Results}
\label{Experimental Results}

The atmospheric neutrino data of the
Super-Kamiokande experiment are well fitted by
$\nu_\mu \to \nu_\tau$
transitions with large mixing \cite{SK-atm-Nu2002}:
\begin{equation}
1.2 \times 10^{-3} \, \mathrm{eV}^2
<
\Delta{m}^2_{\mathrm{atm}}
<
5.0 \times 10^{-3} \, \mathrm{eV}^2
\,,
\quad
\sin^2 2\vartheta_{\mathrm{atm}}
>
0.84
\quad
\text{(99\% C.L.)}
\,,
\label{A}
\end{equation}
where
$\Delta{m}^2_{\mathrm{atm}}$
is the atmospheric neutrino squared-mass difference
and
$\vartheta_{\mathrm{atm}}$
is the effective mixing angles in two-generation analyses
of atmospheric neutrino data.

The global analysis of all solar neutrino data
in terms of
$\nu_e \to \nu_\mu,\nu_\tau$
performed in Ref.~\cite{Bahcall:2002hv}
yielded
\begin{align}
\text{LMA:}
\quad
\null & \null
2.3 \times 10^{-5} \, \mathrm{eV}^2
<
\Delta{m}^2_{\mathrm{sol}}
<
3.7 \times 10^{-4} \, \mathrm{eV}^2
\,,
\quad
0.24
<
\tan^2 \vartheta_{\mathrm{sol}}
<
0.89
\quad
\text{(99.73\% C.L.)}
\,,
\label{LMA}
\\
\text{LOW:}
\quad
\null & \null
3.5 \times 10^{-8} \, \mathrm{eV}^2
<
\Delta{m}^2_{\mathrm{sol}}
<
1.2 \times 10^{-7} \, \mathrm{eV}^2
\,,
\quad
0.43
<
\tan^2 \vartheta_{\mathrm{sol}}
<
0.86
\quad
\text{(99.73\% C.L.)}
\,,
\label{LOW}
\end{align}
where
$\Delta{m}^2_{\mathrm{sol}}$
is the atmospheric neutrino squared-mass difference
and
$\vartheta_{\mathrm{sol}}$
is the effective mixing angles in two-generation analyses
of solar neutrino data.
In Eqs.~(\ref{LMA}) and (\ref{LOW})
we reported only the boundaries of the
so-called LMA and LOW regions,
where matter effects
contribute to neutrino transitions
in the Sun and in the Earth
(see Ref.~\cite{BGG-review-98}).
The LMA region is currently favored,
because it is much larger than the LOW region and
it contains the minimum of the $\chi^2$
(a LOW region appears only at 99\% C.L.).
Additional small VAC regions,
in which only neutrino oscillations in vacuum contribute,
are marginally allowed (at 99.73\% C.L.)
\cite{Bahcall:2002hv}.

The limits in Eqs.~(\ref{LMA}) and (\ref{LOW})
show that also the mixing relevant for solar neutrino oscillations
is large.
However,
maximal mixing seems strongly disfavored
(neglecting the above-mentioned
small and marginal VAC regions)
from the analysis of solar neutrino data
in Ref.~\cite{Bahcall:2002hv}.
This conclusion is supported by the results
of some other authors
\cite{Ahmad:2002ka,Barger:2002iv,deHolanda:2002pp,Fukuda:2002pe},
whereas the authors of
Refs.~\cite{Bandyopadhyay:2002xj,Strumia:2002rv,Fogli:2002pt,Fogli:2002pb}
found slightly larger allowed regions,
with marginally allowed maximal mixing.
Therefore,
it is not clear at present if maximal mixing
in solar neutrino oscillations is excluded or not.
Hopefully,
this problem will be solved soon by the
KamLAND \cite{Dazeley:2002yf}
experiment,
or by the
BOREXINO \cite{Bellini-Nu2002}
experiment.

Solar and atmospheric neutrino data can be well fitted
in the
framework of three-neutrino mixing,
that allows solar
$\nu_e \to \nu_\mu,\nu_\tau$
transitions with
$ \Delta{m}^2_{\mathrm{sol}} = \Delta{m}^2_{21} \equiv m_2^2 - m_1^2 $
and
atmospheric
$\nu_\mu \to \nu_\tau$
transitions with
$ \Delta{m}^2_{\mathrm{atm}} \simeq \Delta{m}^2_{31} \equiv m_3^2 - m_1^2 $
(see Refs.~\cite{Bilenky:1998tw,Giunti-ICHEP02}),
where
$m_1,m_2,m_3$
are the three neutrino masses.

From the results of the CHOOZ long-baseline
reactor neutrino experiment \cite{Apollonio:1999ae},
it is known
that the element $U_{e3}$
of the three-generation neutrino mixing matrix is small \cite{Fogli:2002pb}:
\begin{equation}
|U_{e3}|^2 < 5 \times 10^{-2}
\quad
\text{(99.73\% C.L.)}
\,.
\label{Ue3bound}
\end{equation}
The results of the CHOOZ experiment have been confirmed
by the Palo Verde experiment \cite{Boehm:2001ik},
and by the absence of $\nu_e$ transitions in the
Super-Kamiokande atmospheric neutrino data
\cite{SK-atm-Nu2002}.

An important consequence of the smallness of $U_{e3}$
is the practical decoupling of solar and atmospheric
neutrino oscillations
\cite{Bilenky:1998tw},
which can be analyzed in terms of two-neutrino oscillations
with the effective mixing angles
$\vartheta_{\mathrm{sol}}$
and
$\vartheta_{\mathrm{atm}}$
given by
\begin{align}
\null & \null
\cos^2\vartheta_{\mathrm{sol}}
=
\frac{|U_{e1}|^2}{1-|U_{e3}|^2}
\,,
\quad
\null & \null
\sin^2\vartheta_{\mathrm{sol}}
=
\frac{|U_{e2}|^2}{1-|U_{e3}|^2}
\,,
\label{011}
\\
\null & \null
\cos^2\vartheta_{\mathrm{atm}}
=
\frac{|U_{\tau3}|^2}{1-|U_{e3}|^2}
\,,
\quad
\null & \null
\sin^2\vartheta_{\mathrm{atm}}
=
\frac{|U_{\mu3}|^2}{1-|U_{e3}|^2}
\,.
\label{012}
\end{align}
From the limits in
Eqs.~(\ref{A})--(\ref{Ue3bound})
we get
the following allowed intervals
for the absolute values of the elements
of the mixing matrix
(the intervals are correlated,
because of unitarity):
\begin{equation}
|U|
\approx
\begin{pmatrix}
0.71-0.90 & 0.43-0.69 & 0.00-0.22 \\
0.24-0.66 & 0.40-0.81 & 0.53-0.84 \\
0.24-0.66 & 0.40-0.81 & 0.53-0.84 \\
\end{pmatrix}
\,.
\label{010}
\end{equation}
Hence,
the three-neutrino mixing is bilarge,
not too far from bimaximal
\cite{Vissani:1997pa,Barger:1998ta,Baltz:1998ey}.

As explained in the following section,
in this paper we extend the work presented in
Ref.~\cite{Giunti:2002ye},
in which
we discussed the possibility that
the deviation from bimaximal mixing
is due to the unitary matrix that diagonalizes the
charged lepton mass matrix.

\section{Lepton Mixing}
\label{Lepton Mixing}

Lepton mixing is due to the fact that in general
the charged lepton and neutrino fields in the
weak charged current
\begin{equation}
{j_{\rho}^{\mathrm{CC}}}^\dagger
=
2
\sum_{\alpha'=e',\mu',\tau'}
\overline{\ell_{\alpha'L}}
\gamma_{\rho}
\nu_{\alpha'L}
\label{001}
\end{equation}
do not have a definite mass, but are unitary linear combinations
of massive charged lepton and neutrino fields:
\begin{equation}
\ell_{\alpha'L}
=
\sum_{\alpha=e,\mu,\tau}
V^{(\ell)}_{\alpha'\alpha}
\ell_{\alpha L}
\,,
\qquad
\nu_{\alpha'L}
=
\sum_{k=1}^{3}
V^{(\nu)}_{\alpha'k}
\nu_{kL}
\,.
\label{002}
\end{equation}
The unitary matrices
$V^{(\ell)}$
and
$V^{(\nu)}$
diagonalize, respectively, the charged lepton and neutrino mass matrices.
The weak charged current (\ref{001})
is written in terms of the massive charged lepton and neutrino fields
as
\begin{equation}
{j_{\rho}^{\mathrm{CC}}}^\dagger
=
2
\sum_{\alpha=e,\mu,\tau}
\sum_{k=1}^3
\overline{\ell_{\alpha L}}
\gamma_{\rho}
U_{\alpha k}
\nu_{k L}
\,,
\label{003}
\end{equation}
with the unitary lepton mixing matrix
\begin{equation}
U = {V^{(\ell)}}^{\dagger} V^{(\nu)}
\,.
\label{004}
\end{equation}
Since the charged leptons with definite mass
are directly observable
(through their electromagnetic interactions in detectors),
it is convenient to assign them lepton numbers
$L_{\alpha}$ ($\alpha=e,\mu,\tau$)
and define the corresponding flavor neutrino fields
\begin{equation}
\nu_{\alpha L}
=
\sum_{k=1}^3
U_{\alpha k}
\nu_{k L}
\,.
\label{005}
\end{equation}
In this way,
the weak charged current (\ref{004})
can be written as
\begin{equation}
{j_{\rho}^{\mathrm{CC}}}^\dagger
=
2
\sum_{\alpha=e,\mu,\tau}
\overline{\ell_{\alpha L}}
\gamma_{\rho}
\nu_{\alpha L}
\,,
\label{006}
\end{equation}
showing that the destruction of a flavor neutrino $\nu_\alpha$
(or the creation of a flavor antineutrino $\bar\nu_\alpha$)
is associated with the creation of a charged lepton $\ell_\alpha^-$
(or the destruction of a charged lepton $\ell_\alpha^+$).
However,
as shown in Eq.~(\ref{005}),
a flavor neutrino $\nu_\alpha$
is not an elementary particle,
but the superposition of neutrinos $\nu_k$
with masses $m_k$ ($k=1,2,3$).
This phenomenon is called ``neutrino mixing'' or ``lepton mixing''
and generates neutrino oscillations
(see Ref.~\cite{BGG-review-98}).
The name ``lepton mixing''
appropriately recalls that the mixing matrix $U$ is given by the
product (\ref{002}) of the unitary matrices
$V^{(\ell)}$
and
$V^{(\nu)}$
that diagonalize the charged lepton and neutrino mass matrices.

In Ref.~\cite{Giunti:2002ye}
we supposed that the neutrino unitary matrix $V^{(\nu)}$
has the bimaximal form
\begin{equation}
V^{(\nu)}
=
\begin{pmatrix}
\frac{1}{\sqrt{2}}
&
\frac{1}{\sqrt{2}}
&
0
\\
- \frac{1}{2}
&
\frac{1}{2}
&
\frac{1}{\sqrt{2}}
\\
\frac{1}{2}
&
- \frac{1}{2}
&
\frac{1}{\sqrt{2}}
\end{pmatrix}
\,.
\label{007}
\end{equation}
Although not natural in general,
such bimaximal matrix could be due to an appropriate symmetry
(as a $L_e - L_\mu - L_\tau$ symmetry
\cite{Petcov:1982ya,Barbieri:1998mq,Grimus:2001ex,Babu:2002ex,Goh:2002nk}),
maybe related to the special Majorana nature of neutrinos.
On the other hand,
since
the masses of charged leptons are generated by the
same Higgs mechanism that generates quark masses,
we naturally supposed that
the charged lepton unitary matrix $V^{(\ell)}$
has the CKM form
\begin{equation}
V^{(\ell)}
=
\begin{pmatrix}
c_{12}
c_{13}
&
s_{12}
c_{13}
&
s_{13}
e^{-i\phi_{13}}
\\
-
s_{12}
c_{23}
-
c_{12}
s_{23}
s_{13}
e^{i\phi_{13}}
&
c_{12}
c_{23}
-
s_{12}
s_{23}
s_{13}
e^{i\phi_{13}}
&
s_{23}
c_{13}
\\
s_{12}
s_{23}
-
c_{12}
c_{23}
s_{13}
e^{i\phi_{13}}
&
-
c_{12}
s_{23}
-
s_{12}
c_{23}
s_{13}
e^{i\phi_{13}}
&
c_{23}
c_{13}
\end{pmatrix}
\,,
\label{008}
\end{equation}
where
$s_{ij}\equiv \sin{\vartheta_{ij}}$
and
$c_{ij}\equiv \cos{\vartheta_{ij}}$,
with the hierarchy
\begin{equation}
s_{12} \ll 1
\,,
\qquad
s_{23} \sim s_{12}^2
\,,
\qquad
s_{13} \sim s_{12}^3
\label{009}
\end{equation}
similar to the one in the quark sector
(see Ref.~\cite{PDG}).
Using the matrices (\ref{007}) and (\ref{008})
in Eq.~(\ref{004}),
we found that the size of $U_{e3}$
and
the deviation of the effective solar mixing angle
$\vartheta_{\mathrm{sol}}$
from its bimaximal value $\pi/4$
are related by their leading order proportionality to $s_{12}$.
On the other hand,
the effective atmospheric mixing angle
$\vartheta_{\mathrm{atm}}$
is insensitive
to the contribution of the charged lepton matrix $V^{(\ell)}$,
keeping its bimaximal value
$\pi/4$
up to negligible corrections of order $s_{12}^4$.
This is in agreement with the indications
of maximal mixing found in atmospheric neutrino experiments.
Unfortunately,
the suppositions in Ref.~\cite{Giunti:2002ye}
lead to very small CP violation:
the maximum possible value of the Jarlskog invariant is
more than one order of magnitude smaller than
its absolute maximum possible value.

In this paper we extend the lepton mixing scheme proposed
in Ref.~\cite{Giunti:2002ye}
noting that
in general the unitary matrices $V^{(\ell)}$ and $V^{(\nu)}$
may depend on more phases than the single phase $\phi_{13}$
in Eq.~(\ref{008}).

In the Appendix~\ref{Parameterization and Rephasing of the Mixing Matrix}
we show that in general it is possible to choose
the phases of the charged lepton fields in Eq.~(\ref{003})
in order to write the charged lepton matrix $V^{(\ell)}$
in the form of Eq.~(\ref{008}),
with only one phase $\phi_{13}$
(see Eq.~(\ref{A014})),
but in this case the most general neutrino matrix $V^{(\nu)}$
depends on three angles
$\theta_{12}^{(\nu)}$,
$\theta_{23}^{(\nu)}$,
$\theta_{13}^{(\nu)}$,
three ``Dirac type'' phases
$\psi_{12}$,
$\psi_{23}$,
$\psi_{13}$,
and two ``Majorana type'' phases $\lambda_{21}$, $\lambda_{31}$
(see Eq.~(\ref{A015})).

A bimaximal form for the neutrino matrix $V^{(\nu)}$
is obtained by setting
$ \theta_{12}^{(\nu)} = \theta_{23}^{(\nu)} = \pi/4 $
and
$\theta_{13}^{(\nu)} = 0$,
leading to 
\begin{equation}
V^{(\nu)}
=
W^{(23)}(\pi/4,\psi_{23})
\,
W^{(12)}(\pi/4,\psi_{12})
\,
D(\vec{\lambda})
\,,
\label{022}
\end{equation}
in the notation of
Appendix~\ref{Parameterization and Rephasing of the Mixing Matrix}.
This matrix depends on
two ``Dirac type'' phases
$\psi_{12}$,
$\psi_{23}$.
The diagonal matrix of phases
$D(\vec{\lambda})$
on the right,
with
$\vec{\lambda} = (1,\lambda_{21},\lambda_{31})$,
is present only if neutrinos are Majorana particles.
Explicitly, we have
\begin{equation}
V^{(\nu)}
=
\begin{pmatrix}
e^{i\psi_{12}} & 0 & 0
\\
0 & 1 & 0
\\
0 & 0 & e^{-i\psi_{23}}
\end{pmatrix}
\begin{pmatrix}
\frac{1}{\sqrt{2}}
&
\frac{1}{\sqrt{2}}
&
0
\\
- \frac{1}{2}
&
\frac{1}{2}
&
\frac{1}{\sqrt{2}}
\\
\frac{1}{2}
&
- \frac{1}{2}
&
\frac{1}{\sqrt{2}}
\end{pmatrix}
\begin{pmatrix}
e^{-i\psi_{12}} & 0 & 0
\\
0 & e^{i\lambda_{21}} & 0
\\
0 & 0 & e^{i(\lambda_{31}+\psi_{23})}
\end{pmatrix}
\,.
\label{023}
\end{equation}
In order to write the mixing matrix in this form
we used
the property in Eq.~(\ref{CKM70311}) for the phase $\psi_{12}$
and
the property in Eq.~(\ref{CKM70312}) for the phase $\psi_{23}$.
Obviously,
the matrix
$D^{(1)}\!\left(\psi_{12}\right)$
commutes with
$
W^{(23)}(\pi/4,\psi_{23})
=
{D^{(3)}}^{\dagger}\!\left(\psi_{23}\right)
R^{(23)}(\pi/4)
D^{(3)}\!\left(\psi_{23}\right)
$
and
the matrix
$D^{(3)}\!\left(\psi_{23}\right)$
commutes with
$
W^{(12)}(\pi/4,\psi_{12})
=
D^{(1)}\!\left(\psi_{12}\right)
R^{(12)}(\pi/4)
{D^{(1)}}^{\dagger}\!\left(\psi_{12}\right)
$.

In the following we assume the hierarchy (\ref{009})
for the angles in the charged lepton matrix
and we evaluate all quantities at the leading order
in $s_{12}$.
As a first instance,
the mixing matrix is given by
\begin{equation}
U
=
\begin{pmatrix}
1 & 0 & 0
\\
0 & 1 & 0
\\
0 & 0 & e^{-i\psi_{23}}
\end{pmatrix}
\begin{pmatrix}
\frac{1}{\sqrt{2}} + \frac{1}{2} s_{12} e^{-i\psi_{12}}
&
\frac{1}{\sqrt{2}} e^{i\psi_{12}} - \frac{1}{2} s_{12}
&
- \frac{1}{\sqrt{2}} s_{12}
\\
- \frac{1}{2} e^{-i\psi_{12}} + \frac{1}{\sqrt{2}} s_{12}
&
\frac{1}{2} + \frac{1}{\sqrt{2}} s_{12} e^{i\psi_{12}}
&
\frac{1}{\sqrt{2}}
\\
\frac{1}{2} e^{-i\psi_{12}}
&
- \frac{1}{2}
&
\frac{1}{\sqrt{2}}
\end{pmatrix}
\begin{pmatrix}
1 & 0 & 0
\\
0 & e^{i\lambda_{21}} & 0
\\
0 & 0 & e^{i(\lambda_{31}+\psi_{23})}
\end{pmatrix}
+
\mathrm{O}(s_{12}^2)
\,.
\label{024}
\end{equation}
Therefore,
as in Ref.~\cite{Giunti:2002ye},
the value of $U_{e3}$,
\begin{equation}
U_{e3}
=
- \frac{1}{\sqrt{2}} s_{12} e^{i(\lambda_{31}+\psi_{23})}
+
\mathrm{O}(s_{12}^3)
\,,
\label{Ue3}
\end{equation}
is proportional to $s_{12}$,
at leading order.
From Eq.~(\ref{024})
one can see that at first order in $s_{12}$
the phase $\psi_{23}$ is factorized
in the diagonal matrices on the left and right
of the mixing matrix,
because the matrix
${D^{(3)}}^{\dagger}\!\left(\psi_{23}\right)$
commutes with
$V^{(\ell)}$
at first order in $s_{12}$
($
V^{(\ell)}
=
R^{(12)}\!\left(\vartheta_{12}\right)
+
\mathrm{O}(s_{12}^2)
$).
This implies that,
at first order in $s_{12}$,
the phase
$\psi_{23}$
is irrelevant for neutrino oscillations
in vacuum as well as in matter
(see Ref.~\cite{BGG-review-98}),
which are invariant under the transformations
$
U_{\alpha k}
\to
e^{-i\xi_{\alpha}}
U_{\alpha k}
e^{-i\xi_{k}}
$
with arbitrary phases
$\xi_{\alpha}$ ($\alpha=e,\mu,\tau$)
and
$\xi_{k}$ ($k=1,2,3$)
that can eliminate the phase
$\psi_{23}$
in Eq.~(\ref{024})
(the Majorana phases
$\lambda_{21}$,
$\lambda_{31}$
can be eliminated in any case
and never contribute to neutrino oscillations).

\section{Phenomenology}
\label{Phenomenology}

Since $U_{e3}$ in Eq.~(\ref{Ue3})
is proportional to $s_{12}$,
the value of $|s_{12}|$ is severely limited by
the upper bound for $|U_{e3}|^2$ in Eq.~(\ref{Ue3bound}):
\begin{equation}
|s_{12}| < 0.32
\,.
\label{s12max}
\end{equation}

\subsection{Solar Neutrinos}
\label{Solar Neutrinos}

The effective solar mixing angle
$\vartheta_{\mathrm{sol}}$
is given by
\begin{equation}
\tan^2 \vartheta_{\mathrm{sol}}
=
1
-
2 \sqrt{2} \, s_{12} \cos( \psi_{12} )
+
\mathrm{O}(s_{12}^2)
\,.
\label{t2ts}
\end{equation}
Hence,
at first order in $s_{12}$
the deviation of $\tan^2 \vartheta_{\mathrm{sol}}$
from unity,
which corresponds to maximal mixing,
is not only proportional to
$s_{12}$
as in Ref.~\cite{Giunti:2002ye},
but also to
$\cos( \psi_{12} )$.
This means that in the scheme under consideration
it is possible to
have $U_{e3} \neq 0$
even with maximal solar mixing
(with $\psi_{12}=\pi/2$).
In general,
the contribution of $\cos( \psi_{12} )$ in Eq.~(\ref{t2ts})
allows to have a solar mixing that is maximal or close to maximal.
It is even possible to have an effective mixing angle
$\vartheta_{\mathrm{sol}}$
in the ``dark side''
($ \tan^2 \vartheta_{\mathrm{sol}} > 1 $)
\cite{deGouvea:2000cq}
with negative values of $s_{12} \cos( \psi_{12} )$.

The upper limit on $\tan^2 \vartheta_{\mathrm{sol}}$ in Eq.~(\ref{LMA})
implies that
\begin{equation}
s_{12} \cos( \psi_{12} )
>
0.04
\,.
\label{031}
\end{equation}
This lower limit
is the same as that derived in Ref.~\cite{Giunti:2002ye}
for $s_{12}$ alone,
which follows trivially from Eq.~(\ref{031}):
\begin{equation}
s_{12} > 0.04
\,.
\label{s12min}
\end{equation}
This lower limit for $s_{12}$
leads to the lower bound
\begin{equation}
|U_{e3}| > 0.03
\,,
\label{Ue3min}
\end{equation}
already found in Ref.~\cite{Giunti:2002ye}.
However,
if $ \psi_{12} \neq 0 $,
the lower limit for $s_{12}$ alone may be
significantly larger than 0.04,
leading to a lower bound for $|U_{e3}|$ larger than that in Eq.~(\ref{Ue3min}).
Such values of $|U_{e3}|$
could be measured in the JHF-Kamioka long-baseline neutrino oscillation
experiment,
which has a planned sensitivity of
$|U_{e3}| \simeq 0.04$ at 90\% CL
in the first phase with the Super-Kamiokande detector
and
$|U_{e3}| < 10^{-2}$
in the second phase with the Hyper-Kamiokande detector
\cite{Itow:2001ee}.

From the lower limit for $s_{12} \cos( \psi_{12} )$ in Eq.~(\ref{031})
and the upper bound for $s_{12}$ in Eq.~(\ref{s12max}),
for $\cos( \psi_{12} )$
we get the lower bound
\begin{equation}
\cos( \psi_{12} )
>
0.13
\,.
\label{cos_psi_12_min}
\end{equation}

\begin{figure}
\begin{center}
\includegraphics[bb=0 428 456 751, width=\textwidth]{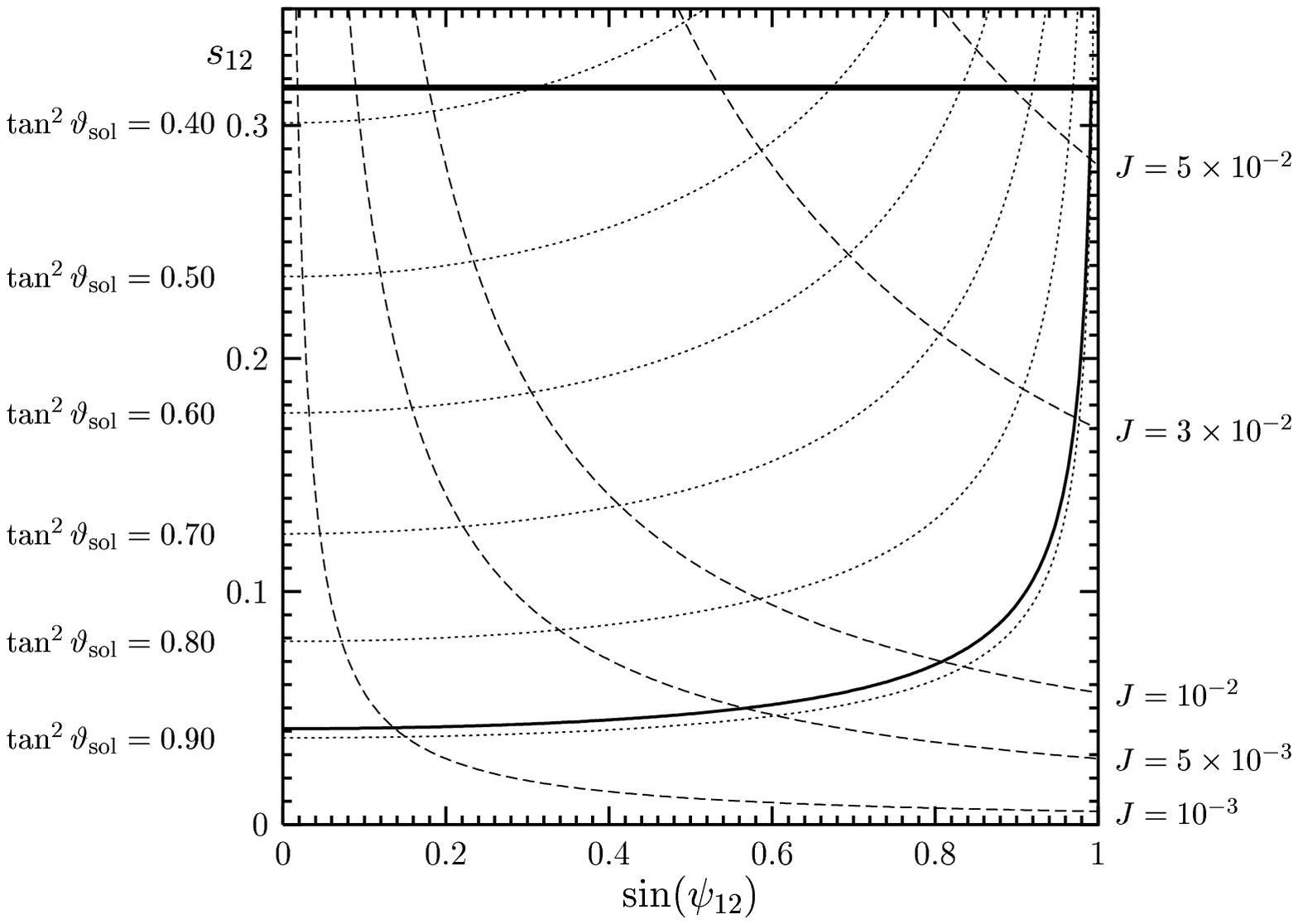}
\end{center}
\caption{ \label{ss}
Allowed region in the positive
$\sin( \psi_{12} )$--$s_{12}$
plane.
The thick solid line represents the upper bound in Eq.~(\ref{s12max}),
$s_{12} < 0.32$.
The thin solid line represent the limit in Eq.~(\ref{031}),
corresponding to the upper limit on $\tan^2 \vartheta_{\mathrm{sol}}$ in
the LMA region (Eq.~(\ref{LMA})).
The dotted lines have
the indicated constant values of
$\tan^2 \vartheta_{\mathrm{sol}}$.
They have been calculated using the exact expression of
$\tan^2 \vartheta_{\mathrm{sol}}$
with
$s_{23} = s_{12}^2$,
$s_{13} = s_{12}^3$,
$\phi_{13} = 0$,
$\psi_{23} = 0$.
The dashed lines have the indicated
constant value of the CP-violation parameter $J$
in Eq.~(\ref{Jdef}).
}
\end{figure}

Figure~\ref{ss}
shows the allowed region in the positive
$\sin( \psi_{12} )$--$s_{12}$
plane,
together with some curves with constant value of
$\tan^2 \vartheta_{\mathrm{sol}}$.
These curves
have been calculated using the exact expression of
$\tan^2 \vartheta_{\mathrm{sol}}$,
because for large values of
$s_{12} \cos( \psi_{12} )$
higher-order terms are not negligible.
In order to perform the calculation,
we assumed, for illustration,
the values
$s_{23} = s_{12}^2$,
$s_{13} = s_{12}^3$,
$\phi_{13} = 0$,
$\psi_{23} = 0$.
Only the curves in the upper-left part of the figure
are modified changing these values,
whereas the thin solid lower-bound curve
is insensitive to the values of
$s_{23}$,
$s_{13}$,
$\phi_{13}$,
$\psi_{23}$,
because the leading order approximation in Eq.~(\ref{t2ts})
is accurate.

\subsection{Atmospheric Neutrinos}
\label{Atmospheric Neutrinos}

As in Ref.~\cite{Giunti:2002ye},
the effective atmospheric mixing angle
is insensitive to $s_{12}$,
remaining practically maximal:
\begin{equation}
\sin^2 2\vartheta_{\mathrm{atm}}
=
1
+
\mathrm{O}(s_{12}^4)
\,.
\label{s2t2a}
\end{equation}
This is consistent with the experimental limit in Eq.~(\ref{A}).

\subsection{Long-Baseline Oscillations and CP Violation}
\label{Long-Baseline Oscillations and CP Violation}

The value of $s_{12}$
could be measured in long-baseline oscillation experiments
\cite{MINOS,ICARUS,Kozlov:2001jv}
sensitive to the largest squared-mass difference
$\Delta{m}^2_{31}$.
In these experiments
the transition probabilities
(neglecting matter effects,
which in any case depend only on
$|U_{e3}|^2$,
$|U_{\mu3}|^2$,
$|U_{\tau3}|^2$
and
$\Delta{m}^2_{31}$)
are well approximated by the standard two-generation formula
with the effective oscillation amplitudes
(see Ref.~\cite{BGG-review-98})
\begin{align}
\null & \null
\sin^2 2 \vartheta_{\nu_e\to\nu_e}^{\mathrm{(LBL)}}
=
4 \, |U_{e3}|^2 \left( 1 - |U_{e3}|^2 \right)
=
2 \, s_{12}^2 + \mathrm{O}(s_{12}^4)
\,,
\label{041}
\\
\null & \null
\sin^2 2 \vartheta_{\nu_\mu\to\nu_e}^{\mathrm{(LBL)}}
=
4 \, |U_{\mu3}|^2 \, |U_{e3}|^2
=
s_{12}^2 + \mathrm{O}(s_{12}^4)
\,,
\label{042}
\end{align}
for $\nu_e\to\nu_e$ and $\nu_\mu\to\nu_e$
or $\bar\nu_\mu\to\bar\nu_e$ transitions,
respectively.
In Eq.~(\ref{042})
we neglected possible CP violation effects
which are measurable only by experiments sensitive
to both the squared-mass differences
$\Delta{m}^2_{31}$
and
$\Delta{m}^2_{21}$.
Indeed,
for vacuum oscillations
\begin{equation}
P_{\nu_\mu\to\nu_e}
-
P_{\bar\nu_\mu\to\bar\nu_e}
=
4 \, J
\left[
\sin\frac{\Delta{m}^2_{21}L}{2E}
+
\sin\frac{\Delta{m}^2_{32}L}{2E}
-
\sin\frac{\Delta{m}^2_{31}L}{2E}
\right]
\,,
\label{043}
\end{equation}
where $L$ is the source-detector distance and $E$
is the neutrino energy.
The Jarlskog parameter
\cite{Jarlskog-PRL-85},
\begin{equation}
J = \mathrm{Im}\left[ U_{e2}^* U_{e3} U_{\mu2} U_{\mu3}^* \right]
\label{Jdef}
\end{equation}
is a measure of CP violation
which is invariant under rephasing of the lepton fields.
In contrast with Ref.~\cite{Giunti:2002ye},
taking into account all the possible phases in the
lepton mixing matrix leads to a linear contribution
of $s_{12}$ to $J$:
\begin{equation}
J
=
\frac{1}{4\sqrt{2}} \, s_{12} \, \sin( \psi_{12} )
+
\mathrm{O}(s_{12}^3)
\,.
\label{J}
\end{equation}
Hence,
if $\psi_{12}$ is not too small,
we expect a sizable CP violation.

The maximum possible value for $|J|$
is obtained for $ \psi_{12} = \pm \pi/2 $.
In this case
Eq.~(\ref{t2ts})
shows that the effective solar mixing is maximal
and independent from the value of $s_{12}$.
The upper limit for $s_{12}$ in Eq.~(\ref{s12max})
leads to
\begin{equation}
|J|_{\mathrm{max}}
\simeq
5 \times 10^{-2}
\,,
\label{Jmax}
\end{equation}
which is not far from the absolute upper limit of the
Jarlskog parameter
(see Ref.~\cite{Dunietz-88})
\begin{equation}
|J|_{\mathrm{max}}^{\mathrm{absolute}}
=
\frac{1}{6\sqrt{3}}
=
9.6 \times 10^{-2}
\,.
\label{Jmax-absolute}
\end{equation}

Although a maximal value of the effective solar mixing angle
may be not completely excluded,
as discussed in Section~\ref{Experimental Results}
it is certainly disfavored by current experimental data
and out of the limits in Eqs.~(\ref{LMA}) and (\ref{LOW}).
Considering the allowed interval
in Eq.~(\ref{LMA}) for $\tan^2 \vartheta_{\mathrm{sol}}$
in the LMA region,
which leads to the lower limit (\ref{031}) for
$s_{12} \cos( \psi_{12} )$,
we have
\begin{equation}
|s_{12} \, \sin( \psi_{12} )|
\leq
\sqrt{
(s_{12}^2)_{\mathrm{max}}
-
( s_{12} \, \sin( \psi_{12} ) )^2_{\mathrm{min}}
}
=
0.31
\,,
\label{s12sp12max}
\end{equation}
leading to a maximal value of $|J|$ practically equal to that
in Eq.~(\ref{Jmax}).
Let us notice that in this case the largeness of
$|J|$ is due to a value of $|s_{12}|$ close to the upper bound
in Eq.~(\ref{s12max})
and a large value of $|\sin( \psi_{12} )|$,
that is however sufficiently different from unity
in order to satisfy the lower limit in Eq.~(\ref{031}).

Figure~\ref{ss}
shows some curves with constant value of
$J$
in the positive
$\sin( \psi_{12} )$--$s_{12}$
plane.
One can see that relatively large values of  $J$,
close to the upper limit in Eq.~(\ref{Jmax}),
can be realized if
$s_{12}$ is not too far from the upper bound in Eq~(\ref{s12max})
and $\psi_{12}$ is not too small.
In this case, CP violation
may be measured in the JHF-Kamioka long-baseline neutrino oscillation
experiment
\cite{Itow:2001ee}
or in a neutrino factory experiment \cite{Albright:2000xi}.

\section{Conclusions}
\label{Conclusions}

We have proposed a scheme of lepton mixing in which
the unitary matrix that diagonalizes
the neutrino mass matrix is bimaximal
and the deviation from bimaximal of the lepton mixing matrix
is due to the unitary matrix that diagonalizes
the charged-lepton mass matrix.
This scheme generalizes the one proposed in
Ref.~\cite{Giunti:2002ye}
by taking into account the possible existence
of additional phases.

The unitary matrix that diagonalizes
the charged-lepton mass matrix is assumed to be hierarchical,
like the quark mixing matrix,
since presumably the
charged-lepton and quark mass matrices
are originated by the same standard Higgs mechanism.
The neutrino mass matrix
is generated by a different mechanism and
the bimaximal form the unitary matrix that diagonalizes
the neutrino mass matrix
could be due to to an appropriate symmetry
and
maybe related to the Majorana nature of neutrinos.

We have shown that in general
it is possible to have a sizable value for $|U_{e3}|$
together with an effective two-neutrino maximal mixing in solar neutrino experiments.
If the effective mixing in solar neutrino experiments
is less than maximal,
as indicated by current data,
$|U_{e3}|$ is bounded from below
(see Eq.~(\ref{Ue3min})).
Such values of $|U_{e3}|$ could be measured
in the JHF-Kamioka long-baseline neutrino oscillation
experiment
\cite{Itow:2001ee}.

In Ref.~\cite{Giunti:2002ye},
it was found that CP violation is small.
Here we have shown that the contribution of the
additional possible phases
allows the violation of CP to be relatively large
(see Eq.~(\ref{Jmax}))
and probably measurable in future experiments
(JHF-Kamioka \cite{Itow:2001ee},
neutrino factory \cite{Albright:2000xi}).

\appendix

\section{Parameterization and Rephasing of the Mixing Matrix}
\label{Parameterization and Rephasing of the Mixing Matrix}

A $ 3 \times 3 $
unitary matrix
$V$
can be written as
(see
\cite{Murnaghan-book-62,Schechter-Valle-COMMENT-80,Schechter-Valle-MASSES-80}
and the appendix of \cite{GKM-atm-98})
\begin{equation}
V
=
\left[
\prod_{a<b}
W^{(ab)}\!\left(\theta_{ab},\eta_{ab}\right)
\right]
D(\vec{\omega})
\qquad
(a,b=1,2,3)
\,,
\label{CKM701}
\end{equation}
with the unitary matrices
\begin{align}
\null & \null
D(\vec{\omega})
=
\mathrm{diag}\!\left(
e^{i\omega_{1}}
\, , \,
e^{i\omega_{2}}
\, , \,
e^{i\omega_{3}}
\right)
\,,
\label{CKM702}
\\
\null & \null
\left[
W^{(ab)}\!\left(\theta_{ab},\eta_{ab}\right)
\right]_{rs}
=
\delta_{rs}
+
\left( c_{ab} - 1 \right)
\left(
\delta_{ra} \, \delta_{sa}
+
\delta_{rb} \, \delta_{sb}
\right)
+
s_{ab}
\left(
e^{i\eta_{ab}} \, \delta_{ra} \, \delta_{sb}
-
e^{-i\eta_{ab}} \, \delta_{rb} \, \delta_{sa}
\right)
\,,
\label{CKM703}
\end{align}
where
$c_{ab}\equiv\cos\theta_{ab}$
and
$s_{ab}\equiv\sin\theta_{ab}$.
Here
$D(\vec{\omega})$
is a diagonal matrix
depending from the set of phases
$\vec{\omega}=(\omega_{1},\omega_{2},\omega_{3})$
and
the matrices
$W^{(ab)}\!\left(\theta_{ab},\eta_{ab}\right)$
are
unitary and unimodular.
For example,
we have
\begin{equation}
W^{(12)}(\theta_{12},\eta_{12})
=
\begin{pmatrix}
\cos\theta_{12} & \sin\theta_{12} \, e^{i\eta_{12}} & 0
\\
-\sin\theta_{12} \, e^{-i\eta_{12}} & \cos\theta_{12} & 0
\\
0 & 0 & 1
\end{pmatrix}
\,.
\label{CKM704}
\end{equation}

With an appropriate choice of the phases
$\omega_k$ and $\eta_{ab}$,
the angles
$\theta_{ab}$
can be limited in the range
\begin{equation}
0 \leq \theta_{ab} \leq \frac{\pi}{2}
\,.
\label{CKM7042}
\end{equation}

The order of the product of the matrices $W^{(ab)}$
in Eq.~(\ref{CKM701})
can be chosen in an arbitrary way.
Different choices of order give different parameterizations.

The matrices
$W^{(ab)}\!\left(\theta_{ab},\eta_{ab}\right)$
satisfy the useful identity\footnote{
Indeed,
\begin{align}
\null & \null
\left[
D(\vec{\xi})
\,
W^{(ab)}\!\left(\theta_{ab},\eta_{ab}\right)
D^{\dagger}(\vec{\xi})
\right]_{rs}
=
\nonumber
\\
\null & \null
\hspace{1cm}
=
\sum_{t,u}
e^{i\xi_{r}} \delta_{rt}
\left[
\delta_{rs}
+
\left( c_{ab} - 1 \right)
\left(
\delta_{ta} \delta_{ua}
+
\delta_{tb} \delta_{ub}
\right)
+
s_{ab}
\left(
e^{i\eta_{ab}} \delta_{ta} \delta_{ub}
-
e^{-i\eta_{ab}} \delta_{tb} \delta_{ua}
\right)
\right]
e^{-i\xi_{s}} \delta_{us}
\nonumber
\\
\null & \null
\hspace{1cm}
=
\delta_{rs}
+
\left( c_{ab} - 1 \right)
\left(
\delta_{ra} \delta_{sa}
+
\delta_{rb} \delta_{sb}
\right)
+
s_{ab}
\left(
e^{i(\eta_{ab}+\xi_{a}-\xi_{b})} \delta_{ra} \delta_{sb}
-
e^{-i(\eta_{ab}+\xi_{a}-\xi_{b})} \delta_{rb} \delta_{sa}
\right)
\nonumber
\\
\null & \null
\hspace{1cm}
=
\left[
W^{(ab)}\!\left(
\theta_{ab},\eta_{ab}+\xi_{a}-\xi_{b}
\right)
\right]_{rs}
\,.
\label{CKM7061}
\end{align}
}
\begin{equation}
D(\vec{\xi})
\,
W^{(ab)}\!\left(\theta_{ab},\eta_{ab}\right)
D^{\dagger}(\vec{\xi})
=
W^{(ab)}\!\left(\theta_{ab},\eta_{ab}+\xi_{a}-\xi_{b}\right)
\,,
\label{CKM706}
\end{equation}
for any choice of the phases $\vec{\xi}=(\xi_1,\xi_2,\xi_3)$.

The identity (\ref{CKM706}) allows to write the matrix
$W^{(ab)}\!\left(\theta_{ab},\eta_{ab}\right)$
as\footnote{
Choosing
$\xi_{a}=-\eta_{ab}$,
$\xi_{b}=0$
in Eq.~(\ref{CKM70311})
and
$\xi_{a}=0$,
$\xi_{b}=\eta_{ab}$
in Eq.~(\ref{CKM70312}).
In both cases
$\xi_{c}=0$
for
$c \neq a,b$.
}
\begin{equation}
W^{(ab)}\!\left(\theta_{ab},\eta_{ab}\right)
=
D^{(a)}\!\left(\eta_{ab}\right)
\,
R^{(ab)}\!\left(\theta_{ab}\right)
\,
{D^{(a)}}^{\dagger}\!\left(\eta_{ab}\right)
\,,
\label{CKM70311}
\end{equation}
or
\begin{equation}
W^{(ab)}\!\left(\theta_{ab},\eta_{ab}\right)
=
{D^{(b)}}^{\dagger}\!\left(\eta_{ab}\right)
\,
R^{(ab)}\!\left(\theta_{ab}\right)
\,
D^{(b)}\!\left(\eta_{ab}\right)
\,,
\label{CKM70312}
\end{equation}
with
\begin{align}
\null & \null
[D^{(a)}\!\left(\eta_{ab}\right)]_{rs}
=
\delta_{rs}
+
\left( e^{i\eta_{ab}} - 1 \right)
\delta_{ra} \, \delta_{sa}
\,,
\label{CKM7032}
\\
\null & \null
[R^{(ab)}\!\left(\theta_{ab}\right)]_{rs}
=
\delta_{rs}
+
\left( \cos\theta_{ab} - 1 \right)
\left( \delta_{ra} \, \delta_{sa} + \delta_{rb} \, \delta_{sb} \right)
+
\sin\theta_{ab}
\left(
\delta_{ra} \, \delta_{sb}
-
\delta_{rb} \, \delta_{sa}
\right)
\,.
\label{CKM7033}
\end{align}
The matrix
$R^{(ab)}\!\left(\theta_{ab}\right)$
operates
a rotation of an angle
$\theta_{ab}$
in the
$a$--$b$
plane.
For example,
we have
\begin{equation}
R^{(12)}\!\left(\theta_{12}\right)
=
\begin{pmatrix}
\cos\theta_{12} & \sin\theta_{12} & 0
\\
-\sin\theta_{12} & \cos\theta_{12} & 0
\\
0 & 0 & 1
\end{pmatrix}
\,,
\qquad
D^{(1)}\!\left(\eta_{12}\right)
=
\begin{pmatrix}
e^{i\eta_{12}} & 0 & 0
\\
0 & 1 & 0
\\
0 & 0 & 1
\end{pmatrix}
\,.
\label{CKM7041}
\end{equation}

Let us consider now the mixing matrix (\ref{004}).
In general,
we can write the unitary matrices
$V^{(\ell)}$
and
$V^{(\nu)}$
using Eq.~(\ref{CKM701}),
leading to
\begin{equation}
U
=
D^{\dagger}(\vec{\omega}^{(\ell)})
\left[
\prod_{a<b}
W^{(ab)}\!\left(\theta^{(\ell)}_{ab},\eta^{(\ell)}_{ab}\right)
\right]^{\dagger}
\left[
\prod_{a<b}
W^{(ab)}\!\left(\theta^{(\nu)}_{ab},\eta^{(\nu)}_{ab}\right)
\right]
D(\vec{\omega}^{(\nu)})
\,,
\label{A001}
\end{equation}
in an obvious notation.
Inserting pairs $D^{\dagger}(\vec{\xi}) D(\vec{\xi})$,
Eq.~(\ref{A001}) can be written as
\begin{equation}
U
=
D^{\dagger}(\vec{\omega}^{(\ell)}+\vec{\xi})
\left[
\prod_{a<b}
D(\vec{\xi})
W^{(ab)}\!\left(\theta^{(\ell)}_{ab},\eta^{(\ell)}_{ab}\right)
D^{\dagger}(\vec{\xi})
\right]^{\dagger}
\left[
\prod_{a<b}
D(\vec{\xi})
W^{(ab)}\!\left(\theta^{(\nu)}_{ab},\eta^{(\nu)}_{ab}\right)
D^{\dagger}(\vec{\xi})
\right]
D(\vec{\omega}^{(\nu)}+\vec{\xi})
\,,
\label{A002}
\end{equation}
and using the identity (\ref{CKM706}) we have
\begin{equation}
U
=
D^{\dagger}(\vec{\omega}^{(\ell)}+\vec{\xi})
\left[
\prod_{a<b}
W^{(ab)}\!\left(\theta^{(\ell)}_{ab},\eta^{(\ell)}_{ab}+\xi_a-\xi_b\right)
\right]^{\dagger}
\left[
\prod_{a<b}
W^{(ab)}\!\left(\theta^{(\nu)}_{ab},\eta^{(\nu)}_{ab}+\xi_a-\xi_b\right)
\right]
D(\vec{\omega}^{(\nu)}+\vec{\xi})
\,.
\label{A003}
\end{equation}

Since there are two independent differences
$\xi_a-\xi_b$,
we can extract two phases from the product of $W$'s.
Let us extract
$\eta^{(\ell)}_{12}$
and
$\eta^{(\ell)}_{23}$
with the choice
\begin{equation}
\xi_1 - \xi_2 = - \eta^{(\ell)}_{12}
\,,
\qquad
\xi_2 - \xi_3 = - \eta^{(\ell)}_{23}
\,.
\label{A004}
\end{equation}
With this choice,
Eq.~(\ref{A003}) can be written as
\begin{align}
U
=
e^{i(\omega_1^{(\nu)}+\xi_1)}
D^{\dagger}(\vec{\omega}^{(\ell)}+\vec{\xi})
\null & \null
\left[
R^{(23)}\!\left(\theta_{23}^{(\ell)}\right)
W^{(13)}\!\left(\theta_{13}^{(\ell)},\phi_{13}\right)
R^{(12)}\!\left(\theta_{12}^{(\ell)}\right)
\right]^{\dagger}
\nonumber
\\
\null & \null
\times
\left[
W^{(23)}\!\left(\theta_{23}^{(\nu)},\psi_{23}\right)
W^{(13)}\!\left(\theta_{13}^{(\nu)},\psi_{13}\right)
W^{(12)}\!\left(\theta_{12}^{(\nu)},\psi_{12}\right)
\right]
D(\vec{\lambda})
\,,
\label{A005}
\end{align}
with
\begin{align}
\null & \null
\phi_{13}
=
\eta^{(\ell)}_{13}+\xi_1-\xi_3
=
\eta^{(\ell)}_{13}-\eta^{(\ell)}_{12}-\eta^{(\ell)}_{23}
\,,
\label{A006}
\\
\null & \null
\psi_{12}
=
\eta^{(\nu)}_{12}+\xi_1-\xi_2
=
\eta^{(\nu)}_{13}-\eta^{(\ell)}_{12}
\,,
\label{A007}
\\
\null & \null
\psi_{23}
=
\eta^{(\nu)}_{23}+\xi_2-\xi_3
=
\eta^{(\nu)}_{13}-\eta^{(\ell)}_{23}
\,,
\label{A008}
\\
\null & \null
\psi_{13}
=
\eta^{(\nu)}_{13}+\xi_1-\xi_3
=
\eta^{(\nu)}_{13}-\eta^{(\ell)}_{12}-\eta^{(\ell)}_{23}
\,,
\label{A009}
\end{align}
and
$\vec{\lambda}=(1,\lambda_{21},\lambda_{31})$,
where
\begin{align}
\null & \null
\lambda_{21}
=
\omega_2^{(\nu)} + \xi_2 - \omega_1^{(\nu)} + \xi_1
=
\omega_2^{(\nu)} - \omega_1^{(\nu)} + \eta^{(\ell)}_{12}
\,,
\label{A011}
\\
\null & \null
\lambda_{31}
=
\omega_3^{(\nu)} + \xi_3 - \omega_1^{(\nu)} + \xi_1
=
\omega_3^{(\nu)} - \omega_1^{(\nu)} + \eta^{(\ell)}_{12} + \eta^{(\ell)}_{23}
\,.
\label{A012}
\end{align}

The overall factor
$e^{i(\omega_1^{(\nu)}+\xi_1)}$
and the diagonal matrix of phases
$D^{\dagger}(\vec{\omega}^{(\ell)}+\vec{\xi})$
on the left of Eq.~(\ref{A005})
can be eliminated
by
appropriate rephasing of
the charged lepton fields in Eq.~(\ref{003}).
On the other hand,
if neutrinos are Majorana particles
the Lagrangian is not invariant under rephasing of the
massive neutrino fields and the
diagonal matrix of phases
$D(\vec{\lambda})$
on the right of Eq.~(\ref{A005})
cannot be eliminated.
Hence,
the physical mixing matrix can be written as
\begin{equation}
U
=
\left[
R^{(23)}\!\left(\vartheta_{23}\right)
W^{(13)}\!\left(\vartheta_{13},\phi_{13}\right)
R^{(12)}\!\left(\vartheta_{12}\right)
\right]^{\dagger}
\left[
W^{(23)}\!\left(\theta_{23}^{(\nu)},\psi_{23}\right)
W^{(13)}\!\left(\theta_{13}^{(\nu)},\psi_{13}\right)
W^{(12)}\!\left(\theta_{12}^{(\nu)},\psi_{12}\right)
\right]
D(\vec{\lambda})
\,,
\label{A013}
\end{equation}
with
$ \vartheta_{ab} = \theta_{ab}^{(\ell)} $.
In other words,
in general the charged lepton and neutrino matrices in Eq.~(\ref{004})
can be written as
\begin{align}
\null & \null
V^{(\ell)}
=
R^{(23)}\!\left(\vartheta_{23}\right)
W^{(13)}\!\left(\vartheta_{13},\phi_{13}\right)
R^{(12)}\!\left(\vartheta_{12}\right)
\,,
\label{A014}
\\
\null & \null
V^{(\nu)}
=
\left[
W^{(23)}\!\left(\theta_{23}^{(\nu)},\psi_{23}\right)
W^{(13)}\!\left(\theta_{13}^{(\nu)},\psi_{13}\right)
W^{(12)}\!\left(\theta_{12}^{(\nu)},\psi_{12}\right)
\right]
D(\vec{\lambda})
\,.
\label{A015}
\end{align}
The charged lepton matrix (\ref{A014})
has the standard explicit form given in Eq.~(\ref{008}),
with three angles $\vartheta_{ab}$ and one phase $\phi_{13}$.
The neutrino matrix (\ref{A015})
depends on three angles $\theta_{ab}^{(\nu)}$,
three ``Dirac type'' phases $\psi_{ab}$
and two ``Majorana type'' phases $\lambda_{21}$, $\lambda_{31}$
(that can be eliminated in the case of Dirac neutrinos).
Notice however,
that Eqs.~(\ref{A007})--(\ref{A012})
show that the phases in the neutrino matrix (\ref{A015})
may be due to the diagonalization of the neutrino mass matrix,
or from the diagonalization of the charged lepton mass matrix,
or both.

For the sake of clarity,
let us finally remark that
the mixing matrix $U$ can be obviously written as
\begin{equation}
U
=
\left[
R^{(23)}\!\left(\tilde\vartheta_{23}\right)
W^{(13)}\!\left(\tilde\vartheta_{13},\tilde\phi_{13}\right)
R^{(12)}\!\left(\tilde\vartheta_{12}\right)
\right]^{\dagger}
D(\vec{\lambda})
\,,
\label{A016}
\end{equation}
in terms of three mixing angles
$\tilde\vartheta_{ab}$
and one physical phase
$\tilde\phi_{13}$.
Our construction leading to Eq.~(\ref{A013})
shows that
in general the mixing angles
$\tilde\vartheta_{ab}$
and the phase
$\tilde\phi_{13}$
depend in a rather complicated way on the angles and phases
of both the matrices
$V^{(\ell)}$
and
$V^{(\nu)}$
that diagonalize, respectively,
the charged lepton and neutrino mass matrices.

\begin{thebibliography}{0}
\expandafter\ifx\csname natexlab\endcsname\relax\def\natexlab#1{#1}\fi
\expandafter\ifx\csname bibnamefont\endcsname\relax
  \def\bibnamefont#1{#1}\fi
\expandafter\ifx\csname bibfnamefont\endcsname\relax
  \def\bibfnamefont#1{#1}\fi
\expandafter\ifx\csname citenamefont\endcsname\relax
  \def\citenamefont#1{#1}\fi
\expandafter\ifx\csname url\endcsname\relax
  \def\url#1{\texttt{#1}}\fi
\expandafter\ifx\csname urlprefix\endcsname\relax\def\urlprefix{URL }\fi
\providecommand{\bibinfo}[2]{#2}
\providecommand{\eprint}[2][]{\url{#2}}

\end{thebibliography}


\begin{thebibliography}{50}
\expandafter\ifx\csname natexlab\endcsname\relax\def\natexlab#1{#1}\fi
\expandafter\ifx\csname bibnamefont\endcsname\relax
  \def\bibnamefont#1{#1}\fi
\expandafter\ifx\csname bibfnamefont\endcsname\relax
  \def\bibfnamefont#1{#1}\fi
\expandafter\ifx\csname citenamefont\endcsname\relax
  \def\citenamefont#1{#1}\fi
\expandafter\ifx\csname url\endcsname\relax
  \def\url#1{\texttt{#1}}\fi
\expandafter\ifx\csname urlprefix\endcsname\relax\def\urlprefix{URL }\fi
\providecommand{\bibinfo}[2]{#2}
\providecommand{\eprint}[2][]{\url{#2}}

\bibitem[{\citenamefont{Fukuda et~al.}(1998)}]{Fukuda:1998mi}
\bibinfo{author}{\bibfnamefont{Y.}~\bibnamefont{Fukuda}} \bibnamefont{et~al.}
  (\bibinfo{collaboration}{Super-Kamiokande}), \bibinfo{journal}{Phys. Rev.
  Lett.} \textbf{\bibinfo{volume}{81}}, \bibinfo{pages}{1562}
  (\bibinfo{year}{1998}), \eprint{hep-ex/9807003}.

\bibitem[{\citenamefont{Ahmad et~al.}(2002{\natexlab{a}})}]{Ahmad:2002jz}
\bibinfo{author}{\bibfnamefont{Q.~R.} \bibnamefont{Ahmad}} \bibnamefont{et~al.}
  (\bibinfo{collaboration}{SNO}), \bibinfo{journal}{Phys. Rev. Lett.}
  \textbf{\bibinfo{volume}{89}}, \bibinfo{pages}{011301}
  (\bibinfo{year}{2002}{\natexlab{a}}),
  \eprint[http://arXiv.org/abs]{nucl-ex/0204008}.

\bibitem[{\citenamefont{Nishikawa}(2002)}]{K2K-Nu2002}
\bibinfo{author}{\bibfnamefont{K.}~\bibnamefont{Nishikawa}}
  (\bibinfo{collaboration}{K2K}), \bibinfo{journal}{Talk at Neutrino 2002,
  25--30 May 2002, Munich, Germany}  (\bibinfo{year}{2002}),
  \urlprefix\url{http://neutrino2002.ph.tum.de/pages/transparencies/nishikawa}.

\bibitem[{\citenamefont{Allison et~al.}(1999)}]{Allison:1999ms}
\bibinfo{author}{\bibfnamefont{W.~W.~M.} \bibnamefont{Allison}}
  \bibnamefont{et~al.} (\bibinfo{collaboration}{Soudan-2}),
  \bibinfo{journal}{Phys. Lett.} \textbf{\bibinfo{volume}{B449}},
  \bibinfo{pages}{137} (\bibinfo{year}{1999}),
  \eprint[http://arXiv.org/abs]{hep-ex/9901024}.

\bibitem[{\citenamefont{Ambrosio et~al.}(2000)}]{Ambrosio:2000qy}
\bibinfo{author}{\bibfnamefont{M.}~\bibnamefont{Ambrosio}} \bibnamefont{et~al.}
  (\bibinfo{collaboration}{MACRO}), \bibinfo{journal}{Phys. Lett.}
  \textbf{\bibinfo{volume}{B478}}, \bibinfo{pages}{5} (\bibinfo{year}{2000}).

\bibitem[{\citenamefont{Cleveland et~al.}(1998)}]{Cleveland:1998nv}
\bibinfo{author}{\bibfnamefont{B.~T.} \bibnamefont{Cleveland}}
  \bibnamefont{et~al.} (\bibinfo{collaboration}{Homestake}),
  \bibinfo{journal}{Astrophys. J.} \textbf{\bibinfo{volume}{496}},
  \bibinfo{pages}{505} (\bibinfo{year}{1998}).

\bibitem[{\citenamefont{Hampel et~al.}(1999)}]{Hampel:1998xg}
\bibinfo{author}{\bibfnamefont{W.}~\bibnamefont{Hampel}} \bibnamefont{et~al.}
  (\bibinfo{collaboration}{GALLEX}), \bibinfo{journal}{Phys. Lett.}
  \textbf{\bibinfo{volume}{B447}}, \bibinfo{pages}{127} (\bibinfo{year}{1999}).

\bibitem[{\citenamefont{Abdurashitov et~al.}(2002)\citenamefont{Abdurashitov,
  N. et~al.}}]{astro-ph/0204245}
\bibinfo{author}{\bibnamefont{Abdurashitov}},
  \bibinfo{author}{\bibfnamefont{J.}~\bibnamefont{N.}}, \bibnamefont{et~al.}
  (\bibinfo{collaboration}{SAGE}), \bibinfo{journal}{J. Exp. Theor. Phys.}
  \textbf{\bibinfo{volume}{95}}, \bibinfo{pages}{181} (\bibinfo{year}{2002}),
  \eprint[http://arXiv.org/abs]{astro-ph/0204245}.

\bibitem[{\citenamefont{Altmann et~al.}(2000)}]{Altmann:2000ft}
\bibinfo{author}{\bibfnamefont{M.}~\bibnamefont{Altmann}} \bibnamefont{et~al.}
  (\bibinfo{collaboration}{GNO}), \bibinfo{journal}{Phys. Lett.}
  \textbf{\bibinfo{volume}{B490}}, \bibinfo{pages}{16} (\bibinfo{year}{2000}),
  \eprint{hep-ex/0006034}.

\bibitem[{\citenamefont{Fukuda et~al.}(2002)}]{Fukuda:2002pe}
\bibinfo{author}{\bibfnamefont{S.}~\bibnamefont{Fukuda}} \bibnamefont{et~al.}
  (\bibinfo{collaboration}{Super-Kamiokande}), \bibinfo{journal}{Phys. Lett.}
  \textbf{\bibinfo{volume}{B539}}, \bibinfo{pages}{179} (\bibinfo{year}{2002}),
  \eprint[http://arXiv.org/abs]{hep-ex/0205075}.

\bibitem[{\citenamefont{Giunti}(2002)}]{Giunti-ICHEP02}
\bibinfo{author}{\bibfnamefont{C.}~\bibnamefont{Giunti}}
  (\bibinfo{year}{2002}), \bibinfo{note}{talk presented at ICHEP02, Amsterdam,
  24-31 July 2002, http://www.ichep02.nl},
  \eprint[http://arXiv.org/abs]{hep-ph/0209103}.

\bibitem[{\citenamefont{Giunti and Laveder}(2002)}]{Neutrino-Unbound}
\bibinfo{author}{\bibfnamefont{C.}~\bibnamefont{Giunti}} \bibnamefont{and}
  \bibinfo{author}{\bibfnamefont{M.}~\bibnamefont{Laveder}}
  (\bibinfo{year}{2002}), \eprint{\textrm{Neutrino Unbound}:
  \texttt{http://www.to.infn.it/\~{}giunti/NU}}.

\bibitem[{\citenamefont{Apollonio et~al.}(1999)}]{Apollonio:1999ae}
\bibinfo{author}{\bibfnamefont{M.}~\bibnamefont{Apollonio}}
  \bibnamefont{et~al.} (\bibinfo{collaboration}{CHOOZ}),
  \bibinfo{journal}{Phys. Lett.} \textbf{\bibinfo{volume}{B466}},
  \bibinfo{pages}{415} (\bibinfo{year}{1999}),
  \eprint[http://arXiv.org/abs]{hep-ex/9907037}.

\bibitem[{\citenamefont{Vissani}(1997)}]{Vissani:1997pa}
\bibinfo{author}{\bibfnamefont{F.}~\bibnamefont{Vissani}}
  (\bibinfo{year}{1997}), \eprint[http://arXiv.org/abs]{hep-ph/9708483}.

\bibitem[{\citenamefont{Barger et~al.}(1998)\citenamefont{Barger, Pakvasa,
  Weiler, and Whisnant}}]{Barger:1998ta}
\bibinfo{author}{\bibfnamefont{V.}~\bibnamefont{Barger}},
  \bibinfo{author}{\bibfnamefont{S.}~\bibnamefont{Pakvasa}},
  \bibinfo{author}{\bibfnamefont{T.~J.} \bibnamefont{Weiler}},
  \bibnamefont{and} \bibinfo{author}{\bibfnamefont{K.}~\bibnamefont{Whisnant}},
  \bibinfo{journal}{Phys. Lett.} \textbf{\bibinfo{volume}{B437}},
  \bibinfo{pages}{107} (\bibinfo{year}{1998}), \eprint{hep-ph/9806387}.

\bibitem[{\citenamefont{Baltz et~al.}(1998)\citenamefont{Baltz, Goldhaber, and
  Goldhaber}}]{Baltz:1998ey}
\bibinfo{author}{\bibfnamefont{A.~J.} \bibnamefont{Baltz}},
  \bibinfo{author}{\bibfnamefont{A.~S.} \bibnamefont{Goldhaber}},
  \bibnamefont{and}
  \bibinfo{author}{\bibfnamefont{M.}~\bibnamefont{Goldhaber}},
  \bibinfo{journal}{Phys. Rev. Lett.} \textbf{\bibinfo{volume}{81}},
  \bibinfo{pages}{5730} (\bibinfo{year}{1998}),
  \eprint[http://arXiv.org/abs]{hep-ph/9806540}.

\bibitem[{\citenamefont{Giunti and Tanimoto}(2002)}]{Giunti:2002ye}
\bibinfo{author}{\bibfnamefont{C.}~\bibnamefont{Giunti}} \bibnamefont{and}
  \bibinfo{author}{\bibfnamefont{M.}~\bibnamefont{Tanimoto}},
  \bibinfo{journal}{Phys. Rev.} \textbf{\bibinfo{volume}{D66}},
  \bibinfo{pages}{053013} (\bibinfo{year}{2002}),
  \eprint[http://arXiv.org/abs]{hep-ph/0207096}.

\bibitem[{\citenamefont{Xing}(2001)}]{Xing:2001cx}
\bibinfo{author}{\bibfnamefont{Z.~Z.} \bibnamefont{Xing}},
  \bibinfo{journal}{Phys. Rev.} \textbf{\bibinfo{volume}{D64}},
  \bibinfo{pages}{093013} (\bibinfo{year}{2001}),
  \eprint[http://arXiv.org/abs]{hep-ph/0107005}.

\bibitem[{\citenamefont{Shiozawa}(2002)}]{SK-atm-Nu2002}
\bibinfo{author}{\bibfnamefont{M.}~\bibnamefont{Shiozawa}}
  (\bibinfo{collaboration}{Super-Kamiokande}), \bibinfo{journal}{Talk at
  Neutrino 2002, 25--30 May 2002, Munich, Germany}  (\bibinfo{year}{2002}),
  \urlprefix\url{http://neutrino2002.ph.tum.de/pages/transparencies/shiozawa}.

\bibitem[{\citenamefont{Bahcall et~al.}(2002)\citenamefont{Bahcall,
  Gonzalez-Garcia, and Pena-Garay}}]{Bahcall:2002hv}
\bibinfo{author}{\bibfnamefont{J.~N.} \bibnamefont{Bahcall}},
  \bibinfo{author}{\bibfnamefont{M.~C.} \bibnamefont{Gonzalez-Garcia}},
  \bibnamefont{and}
  \bibinfo{author}{\bibfnamefont{C.}~\bibnamefont{Pena-Garay}},
  \bibinfo{journal}{JHEP} \textbf{\bibinfo{volume}{07}}, \bibinfo{pages}{054}
  (\bibinfo{year}{2002}), \eprint[http://arXiv.org/abs]{hep-ph/0204314}.

\bibitem[{\citenamefont{Bilenky et~al.}(1999)\citenamefont{Bilenky, Giunti, and
  Grimus}}]{BGG-review-98}
\bibinfo{author}{\bibfnamefont{S.~M.} \bibnamefont{Bilenky}},
  \bibinfo{author}{\bibfnamefont{C.}~\bibnamefont{Giunti}}, \bibnamefont{and}
  \bibinfo{author}{\bibfnamefont{W.}~\bibnamefont{Grimus}},
  \bibinfo{journal}{Prog. Part. Nucl. Phys.} \textbf{\bibinfo{volume}{43}},
  \bibinfo{pages}{1} (\bibinfo{year}{1999}), \eprint{hep-ph/9812360}.

\bibitem[{\citenamefont{Ahmad et~al.}(2002{\natexlab{b}})}]{Ahmad:2002ka}
\bibinfo{author}{\bibfnamefont{Q.~R.} \bibnamefont{Ahmad}} \bibnamefont{et~al.}
  (\bibinfo{collaboration}{SNO}), \bibinfo{journal}{Phys. Rev. Lett.}
  \textbf{\bibinfo{volume}{89}}, \bibinfo{pages}{011302}
  (\bibinfo{year}{2002}{\natexlab{b}}),
  \eprint[http://arXiv.org/abs]{nucl-ex/0204009}.

\bibitem[{\citenamefont{Barger et~al.}(2002)\citenamefont{Barger, Marfatia,
  Whisnant, and Wood}}]{Barger:2002iv}
\bibinfo{author}{\bibfnamefont{V.}~\bibnamefont{Barger}},
  \bibinfo{author}{\bibfnamefont{D.}~\bibnamefont{Marfatia}},
  \bibinfo{author}{\bibfnamefont{K.}~\bibnamefont{Whisnant}}, \bibnamefont{and}
  \bibinfo{author}{\bibfnamefont{B.~P.} \bibnamefont{Wood}},
  \bibinfo{journal}{Phys. Lett.} \textbf{\bibinfo{volume}{B537}},
  \bibinfo{pages}{179} (\bibinfo{year}{2002}),
  \eprint[http://arXiv.org/abs]{hep-ph/0204253}.

\bibitem[{\citenamefont{de~Holanda and Smirnov}(2002)}]{deHolanda:2002pp}
\bibinfo{author}{\bibfnamefont{P.~C.} \bibnamefont{de~Holanda}}
  \bibnamefont{and} \bibinfo{author}{\bibfnamefont{A.~Y.}
  \bibnamefont{Smirnov}} (\bibinfo{year}{2002}),
  \eprint[http://arXiv.org/abs]{hep-ph/0205241}.

\bibitem[{\citenamefont{Bandyopadhyay et~al.}(2002)\citenamefont{Bandyopadhyay,
  Choubey, Goswami, and Roy}}]{Bandyopadhyay:2002xj}
\bibinfo{author}{\bibfnamefont{A.}~\bibnamefont{Bandyopadhyay}},
  \bibinfo{author}{\bibfnamefont{S.}~\bibnamefont{Choubey}},
  \bibinfo{author}{\bibfnamefont{S.}~\bibnamefont{Goswami}}, \bibnamefont{and}
  \bibinfo{author}{\bibfnamefont{D.~P.} \bibnamefont{Roy}},
  \bibinfo{journal}{Phys. Lett.} \textbf{\bibinfo{volume}{B540}},
  \bibinfo{pages}{14} (\bibinfo{year}{2002}),
  \eprint[http://arXiv.org/abs]{hep-ph/0204286}.

\bibitem[{\citenamefont{Strumia et~al.}(2002)\citenamefont{Strumia, Cattadori,
  Ferrari, and Vissani}}]{Strumia:2002rv}
\bibinfo{author}{\bibfnamefont{A.}~\bibnamefont{Strumia}},
  \bibinfo{author}{\bibfnamefont{C.}~\bibnamefont{Cattadori}},
  \bibinfo{author}{\bibfnamefont{N.}~\bibnamefont{Ferrari}}, \bibnamefont{and}
  \bibinfo{author}{\bibfnamefont{F.}~\bibnamefont{Vissani}}
  (\bibinfo{year}{2002}), \eprint[http://arXiv.org/abs]{hep-ph/0205261}.

\bibitem[{\citenamefont{Fogli et~al.}(2002{\natexlab{a}})\citenamefont{Fogli,
  Lisi, Marrone, Montanino, and Palazzo}}]{Fogli:2002pt}
\bibinfo{author}{\bibfnamefont{G.~L.} \bibnamefont{Fogli}},
  \bibinfo{author}{\bibfnamefont{E.}~\bibnamefont{Lisi}},
  \bibinfo{author}{\bibfnamefont{A.}~\bibnamefont{Marrone}},
  \bibinfo{author}{\bibfnamefont{D.}~\bibnamefont{Montanino}},
  \bibnamefont{and} \bibinfo{author}{\bibfnamefont{A.}~\bibnamefont{Palazzo}},
  \bibinfo{journal}{Phys. Rev.} \textbf{\bibinfo{volume}{D66}},
  \bibinfo{pages}{053010} (\bibinfo{year}{2002}{\natexlab{a}}),
  \eprint[http://arXiv.org/abs]{hep-ph/0206162}.

\bibitem[{\citenamefont{Fogli et~al.}(2002{\natexlab{b}})}]{Fogli:2002pb}
\bibinfo{author}{\bibfnamefont{G.~L.} \bibnamefont{Fogli}} \bibnamefont{et~al.}
  (\bibinfo{year}{2002}{\natexlab{b}}),
  \eprint[http://arXiv.org/abs]{hep-ph/0208026}.

\bibitem[{\citenamefont{Dazeley}(2002)}]{Dazeley:2002yf}
\bibinfo{author}{\bibfnamefont{S.~A.} \bibnamefont{Dazeley}}
  (\bibinfo{collaboration}{KamLAND}) (\bibinfo{year}{2002}),
  \eprint[http://arXiv.org/abs]{hep-ex/0205041}.

\bibitem[{\citenamefont{Bellini}(2002)}]{Bellini-Nu2002}
\bibinfo{author}{\bibfnamefont{G.}~\bibnamefont{Bellini}}
  (\bibinfo{collaboration}{BOREXINO}), \bibinfo{journal}{Talk at Neutrino 2002,
  25--30 May 2002, Munich, Germany}  (\bibinfo{year}{2002}),
  \urlprefix\url{http://neutrino2002.ph.tum.de/pages/transparencies/bellini}.

\bibitem[{\citenamefont{Bilenky and Giunti}(1998)}]{Bilenky:1998tw}
\bibinfo{author}{\bibfnamefont{S.~M.} \bibnamefont{Bilenky}} \bibnamefont{and}
  \bibinfo{author}{\bibfnamefont{C.}~\bibnamefont{Giunti}},
  \bibinfo{journal}{Phys. Lett.} \textbf{\bibinfo{volume}{B444}},
  \bibinfo{pages}{379} (\bibinfo{year}{1998}),
  \eprint[http://arXiv.org/abs]{hep-ex/0206047}.

\bibitem[{\citenamefont{Boehm et~al.}(2001)}]{Boehm:2001ik}
\bibinfo{author}{\bibfnamefont{F.}~\bibnamefont{Boehm}} \bibnamefont{et~al.},
  \bibinfo{journal}{Phys. Rev.} \textbf{\bibinfo{volume}{D64}},
  \bibinfo{pages}{112001} (\bibinfo{year}{2001}),
  \eprint[http://arXiv.org/abs]{hep-ex/0107009}.

\bibitem[{\citenamefont{Petcov}(1982)}]{Petcov:1982ya}
\bibinfo{author}{\bibfnamefont{S.~T.} \bibnamefont{Petcov}},
  \bibinfo{journal}{Phys. Lett.} \textbf{\bibinfo{volume}{B110}},
  \bibinfo{pages}{245} (\bibinfo{year}{1982}).

\bibitem[{\citenamefont{Barbieri et~al.}(1998)\citenamefont{Barbieri, Hall,
  Smith, Strumia, and Weiner}}]{Barbieri:1998mq}
\bibinfo{author}{\bibfnamefont{R.}~\bibnamefont{Barbieri}},
  \bibinfo{author}{\bibfnamefont{L.~J.} \bibnamefont{Hall}},
  \bibinfo{author}{\bibfnamefont{D.~R.} \bibnamefont{Smith}},
  \bibinfo{author}{\bibfnamefont{A.}~\bibnamefont{Strumia}}, \bibnamefont{and}
  \bibinfo{author}{\bibfnamefont{N.}~\bibnamefont{Weiner}},
  \bibinfo{journal}{JHEP} \textbf{\bibinfo{volume}{12}}, \bibinfo{pages}{017}
  (\bibinfo{year}{1998}), \eprint[http://arXiv.org/abs]{hep-ph/9807235}.

\bibitem[{\citenamefont{Babu and Mohapatra}(2002)}]{Babu:2002ex}
\bibinfo{author}{\bibfnamefont{K.~S.} \bibnamefont{Babu}} \bibnamefont{and}
  \bibinfo{author}{\bibfnamefont{R.~N.} \bibnamefont{Mohapatra}},
  \bibinfo{journal}{Phys. Lett.} \textbf{\bibinfo{volume}{B532}},
  \bibinfo{pages}{77} (\bibinfo{year}{2002}),
  \eprint[http://arXiv.org/abs]{hep-ph/0201176}.

\bibitem[{\citenamefont{Goh et~al.}(2002)\citenamefont{Goh, Mohapatra, and
  Ng}}]{Goh:2002nk}
\bibinfo{author}{\bibfnamefont{H.~S.} \bibnamefont{Goh}},
  \bibinfo{author}{\bibfnamefont{R.~N.} \bibnamefont{Mohapatra}},
  \bibnamefont{and} \bibinfo{author}{\bibfnamefont{S.~P.} \bibnamefont{Ng}},
  \bibinfo{journal}{Phys. Lett.} \textbf{\bibinfo{volume}{B542}},
  \bibinfo{pages}{116} (\bibinfo{year}{2002}),
  \eprint[http://arXiv.org/abs]{hep-ph/0205131}.

\bibitem[{\citenamefont{Grimus and Lavoura}(2001)}]{Grimus:2001ex}
\bibinfo{author}{\bibfnamefont{W.}~\bibnamefont{Grimus}} \bibnamefont{and}
  \bibinfo{author}{\bibfnamefont{L.}~\bibnamefont{Lavoura}},
  \bibinfo{journal}{JHEP} \textbf{\bibinfo{volume}{07}}, \bibinfo{pages}{045}
  (\bibinfo{year}{2001}), \eprint[http://arXiv.org/abs]{hep-ph/0105212}.

\bibitem[{\citenamefont{Groom et~al.}(2000)}]{PDG}
\bibinfo{author}{\bibfnamefont{D.~E.} \bibnamefont{Groom}} \bibnamefont{et~al.}
  (\bibinfo{collaboration}{Particle Data Group}), \bibinfo{journal}{Eur. Phys.
  J.} \textbf{\bibinfo{volume}{C15}}, \bibinfo{pages}{1}
  (\bibinfo{year}{2000}), \urlprefix\url{http://pdg.lbl.gov}.

\bibitem[{\citenamefont{de~Gouvea et~al.}(2000)\citenamefont{de~Gouvea,
  Friedland, and Murayama}}]{deGouvea:2000cq}
\bibinfo{author}{\bibfnamefont{A.}~\bibnamefont{de~Gouvea}},
  \bibinfo{author}{\bibfnamefont{A.}~\bibnamefont{Friedland}},
  \bibnamefont{and} \bibinfo{author}{\bibfnamefont{H.}~\bibnamefont{Murayama}},
  \bibinfo{journal}{Phys. Lett.} \textbf{\bibinfo{volume}{B490}},
  \bibinfo{pages}{125} (\bibinfo{year}{2000}), \eprint{hep-ph/0002064}.

\bibitem[{\citenamefont{Itow et~al.}(2001)}]{Itow:2001ee}
\bibinfo{author}{\bibfnamefont{Y.}~\bibnamefont{Itow}} \bibnamefont{et~al.}
  (\bibinfo{year}{2001}), \eprint[http://arXiv.org/abs]{hep-ex/0106019}.

\bibitem[{\citenamefont{Adamson et~al.}(1999)}]{MINOS}
\bibinfo{author}{\bibfnamefont{P.}~\bibnamefont{Adamson}} \bibnamefont{et~al.}
  (\bibinfo{collaboration}{MINOS}), \bibinfo{journal}{NuMI-L-476}
  (\bibinfo{year}{1999}),
  \urlprefix\url{http://www.hep.anl.gov/ndk/hypertext/numi.html}.

\bibitem[{\citenamefont{Cennini et~al.}(1994)}]{ICARUS}
\bibinfo{author}{\bibfnamefont{P.}~\bibnamefont{Cennini}} \bibnamefont{et~al.}
  (\bibinfo{collaboration}{ICARUS}), \bibinfo{journal}{LNGS-94/99-I}
  (\bibinfo{year}{1994}), \urlprefix\url{http://pcnometh4.cern.ch}.

\bibitem[{\citenamefont{Kozlov et~al.}(2001)\citenamefont{Kozlov, Mikaelyan,
  and Sinev}}]{Kozlov:2001jv}
\bibinfo{author}{\bibfnamefont{Y.}~\bibnamefont{Kozlov}},
  \bibinfo{author}{\bibfnamefont{L.}~\bibnamefont{Mikaelyan}},
  \bibnamefont{and} \bibinfo{author}{\bibfnamefont{V.}~\bibnamefont{Sinev}}
  (\bibinfo{year}{2001}), \eprint{arXiv:hep-ph/0109277}.

\bibitem[{\citenamefont{Jarlskog}(1985)}]{Jarlskog-PRL-85}
\bibinfo{author}{\bibfnamefont{C.}~\bibnamefont{Jarlskog}},
  \bibinfo{journal}{Phys. Rev. Lett.} \textbf{\bibinfo{volume}{55}},
  \bibinfo{pages}{1039} (\bibinfo{year}{1985}).

\bibitem[{\citenamefont{Dunietz}(1988)}]{Dunietz-88}
\bibinfo{author}{\bibfnamefont{I.}~\bibnamefont{Dunietz}},
  \bibinfo{journal}{Ann. Phys.} \textbf{\bibinfo{volume}{184}},
  \bibinfo{pages}{350} (\bibinfo{year}{1988}).

\bibitem[{\citenamefont{Albright et~al.}(2000)}]{Albright:2000xi}
\bibinfo{author}{\bibfnamefont{C.}~\bibnamefont{Albright}} \bibnamefont{et~al.}
  (\bibinfo{year}{2000}), \eprint{hep-ex/0008064}.

\bibitem[{\citenamefont{Murnaghan}(1962)}]{Murnaghan-book-62}
\bibinfo{author}{\bibfnamefont{F.~D.} \bibnamefont{Murnaghan}},
  \emph{\bibinfo{title}{The unitary and rotation groups}}
  (\bibinfo{publisher}{Spartan Books}, \bibinfo{address}{Washington D.C.},
  \bibinfo{year}{1962}).

\bibitem[{\citenamefont{Schechter and
  Valle}(1980{\natexlab{a}})}]{Schechter-Valle-COMMENT-80}
\bibinfo{author}{\bibfnamefont{J.}~\bibnamefont{Schechter}} \bibnamefont{and}
  \bibinfo{author}{\bibfnamefont{J.~W.~F.} \bibnamefont{Valle}},
  \bibinfo{journal}{Phys. Rev.} \textbf{\bibinfo{volume}{D21}},
  \bibinfo{pages}{309} (\bibinfo{year}{1980}{\natexlab{a}}).

\bibitem[{\citenamefont{Schechter and
  Valle}(1980{\natexlab{b}})}]{Schechter-Valle-MASSES-80}
\bibinfo{author}{\bibfnamefont{J.}~\bibnamefont{Schechter}} \bibnamefont{and}
  \bibinfo{author}{\bibfnamefont{J.~W.~F.} \bibnamefont{Valle}},
  \bibinfo{journal}{Phys. Rev.} \textbf{\bibinfo{volume}{D22}},
  \bibinfo{pages}{2227} (\bibinfo{year}{1980}{\natexlab{b}}).

\bibitem[{\citenamefont{Giunti et~al.}(1998)\citenamefont{Giunti, Kim, and
  Monteno}}]{GKM-atm-98}
\bibinfo{author}{\bibfnamefont{C.}~\bibnamefont{Giunti}},
  \bibinfo{author}{\bibfnamefont{C.~W.} \bibnamefont{Kim}}, \bibnamefont{and}
  \bibinfo{author}{\bibfnamefont{M.}~\bibnamefont{Monteno}},
  \bibinfo{journal}{Nucl. Phys.} \textbf{\bibinfo{volume}{B521}},
  \bibinfo{pages}{3} (\bibinfo{year}{1998}), \eprint{hep-ph/9709439}.

\end{thebibliography}
\end{document}